\documentclass{aastex631}
\usepackage{graphicx} 
\usepackage{gensymb}
\usepackage{xcolor}
\usepackage{soul}
\usepackage{CJK}
\usepackage{hyperref}
\usepackage{tikz}

\shorttitle{GCR depression in 2017}

\shortauthors{Aslam et al.}
\graphicspath{{./}{figures/}}

\begin{document}
\begin{CJK*}{UTF8}{gbsn}

\title{The source of the 2017 cosmic ray half-year modulation event}

\author[0000-0001-9521-3874]{O.P.M. Aslam}
\affiliation{School of Mathematics and Statistics, University of Glasgow, Glasgow G12 8QQ, United Kingdom}

\author[0000-0003-2297-9312]{D. MacTaggart}
\affiliation{School of Mathematics and Statistics,
University of Glasgow,  Glasgow G12 8QQ, United Kingdom}

\author[0000-0002-5808-7239]{R. Battiston}
\affiliation{Department of Physics, University of Trento, Povo, Italy}

\author[0000-0003-0793-7333]{M.S. Potgieter}
\affiliation{Institute for Experimental and Applied Physics (IEAP), Christian-Albrechts-University of Kiel, 24118 Kiel, Germany}

\author[0000-0001-5844-3419]{M.D. Ngobeni}
\affiliation{Centre for Space Research, North-West University,
2520 Potchefstroom, South Africa}

 
\begin{abstract}
In 2017, as the solar cycle approached solar minimum, an unusually long and large depression
was observed in galactic cosmic ray (GCR) protons, detected with the Alpha Magnetic Spectrometer (AMS-02), lasting for the second half of that year. The depression, as seen in the Bartel rotation-averaged proton flux, has the form of a Forbush decrease (FD). Despite this resemblance, however, the cause of the observed depression does not have such a simple explanation as FDs, due to coronal mass ejections (CMEs), typically last for a few days at 1 AU rather than half a year. In this work, we seek the cause of the observed depression and investigate two main possibilities. First, we consider a minicycle - a temporary change in the solar dynamo that changes the behaviour of the global solar magnetic field and, by this, the modulation of GCRs. Secondly, we investigate the behaviour of solar activity, both CMEs and corotating/stream interactions regions (C/SIRs), during this period. Our findings show that, although there is some evidence for minicycle behaviour prior to the depression, the depression is ultimately due to a combination of recurrent CMEs, SIRs and CIRs. A particular characteristic of the depression is that the largest impacts that help to create and maintain it are due to four CMEs from the same, highly active, magnetic source that persists for several solar rotations. This active magnetic source is unusual given the closeness of the solar cycle to solar minimum, which also helps to make the depression more evident.
 
\end{abstract}

\section{Introduction} \label{sec:intro}

Galactic cosmic rays (GCRs) are high energy particles of galactic origin that become modulated inside the heliosphere by the solar wind and the heliospheric magnetic field (HMF). 
The cyclic nature of the Sun's global magnetic field together with transient solar eruptions are responsible for a time-dependent variation in GCRs, at the Earth, with particle rigidities below $\sim$30 GV \citep{2014AdSpR..53.1015S}. The large-scale modulation pattern of GCRs affected by the Sun follows closely the 11-year solar cycle \citep{2013SSRv..176....3C, 2023LRSP...20....2U}. For shorter-term fluctuations to GCR fluxes in the heliosphere, one important influence is \emph{solar activity}, i.e. magnetic phenomena in interplanetary space whose origin is the Sun. Typically, such fluctuations range from the order of a few days to the period of a solar rotation \citep{2000SSRv...93...55C}. 

\begin{figure}[ht!]
\centering
\includegraphics[width=0.9\textwidth]{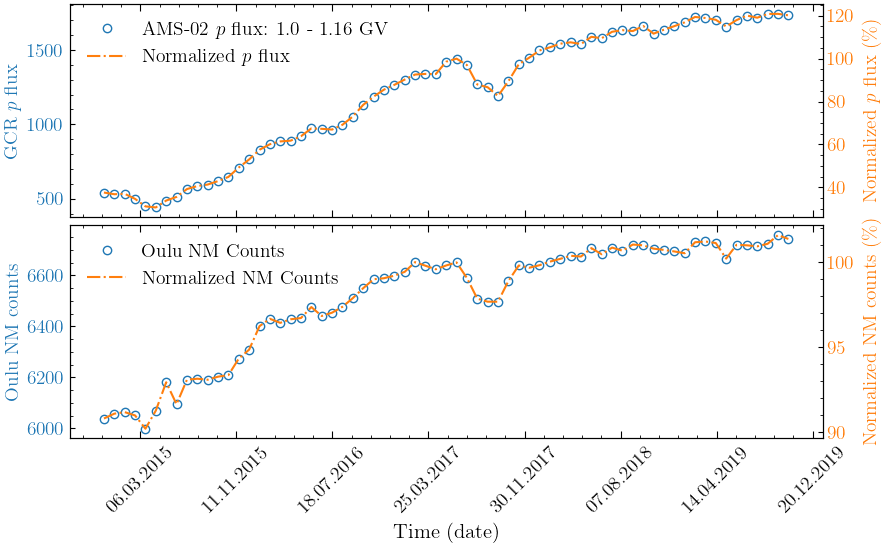}
\caption{Top: Bartel rotations-averaged galactic proton flux variation over the rigidity 1.0 -1.16 GV from 01 December 2014 (BR2474) to 29 October 2019 (BR2540), reported by \cite{2021PhRvL.127A1102A}. Bottom: Same resolution neutron monitor counts recorded by Oulu neutron monitor for the same time period. Left scale shows the flux or NM counts, and the right scale indicates the normalized flux or counts with respect to BR2508.} 
\label{fig1}
\end{figure}

\cite{2021PhRvL.127A1102A} report GCR proton observations measured by the \emph{Alpha Magnetic Spectrometer 02} \citep[AMS-02;][]{2008NIMPA.588..227B,2013NuPhS.243...12T}, located on the International Space Station, that show a large and long depression, between 3-4 Bartel rotations (BRs), at lower rigidity, during the second half of 2017. This depression spans about one fifth of the overall solar cycle variation. Such a large variation was unexpected at that time as solar activity was progressing towards the minimum of Solar Cycle 24. In Figure \ref{fig1}, the top panel shows the variation of Bartel rotations-averaged galactic proton observations of rigidity 1.0 - 1.16 GV, from BRs 2474 - 2540, 01 December 2014 - 29 October 2019, reported by \cite{2021PhRvL.127A1102A}. The left scale indicates the proton flux and the right scale is the normalized flux with respect to BR 2508, 06 June - 03 July 2017, when the depression initiates. The bottom panel of Figure \ref{fig1} shows the variation of neutron monitor (NM) counts recorded by the \href{https://cosmicrays.oulu.fi/}{Oulu NM} for the same period of BRs 2474 - 2540. The left scale indicates the NM counts per minute and the right scale displays the normalized counts with respect to BR 2508. 
The AMS-02 measurements of GCR proton flux show a decrease of 17.43\% within 4 BRs, which, as mentioned above, is almost one fifth of the overall variation (90.22\%) reported from the December 2014 - October 2019 period, in the rigidity bin 1.0 - 1.16 GV. This variation is mirrored in the depression observed in the NM counts, where although the decrease is 2.32\%, this represents $\sim$20\% of the overall variation from December 2014 - October 2019 (and 11.41\% from solar maximum to  minimum).

\begin{figure*}
\gridline{\fig{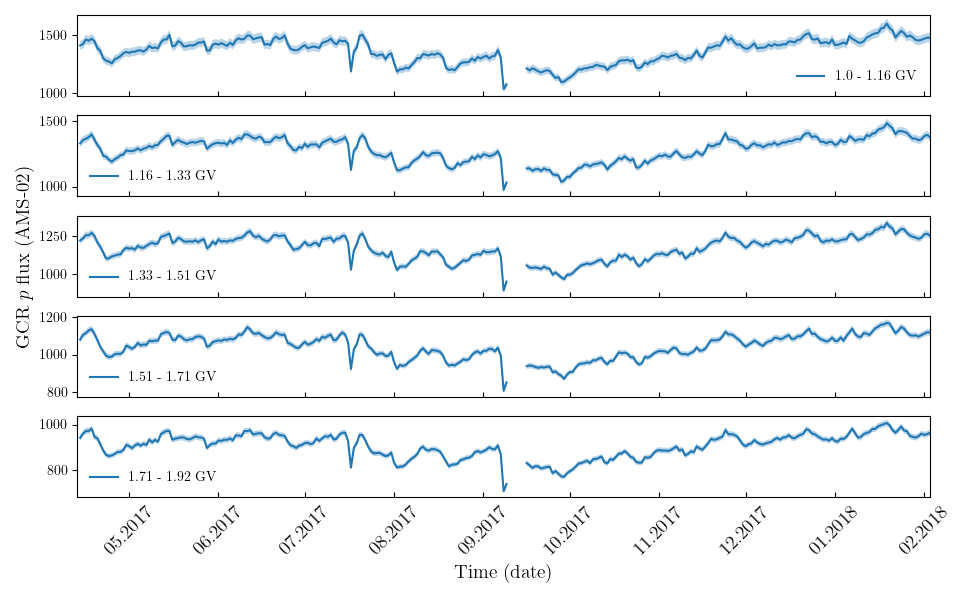}{0.80\textwidth}{(a)}}
\gridline{\fig{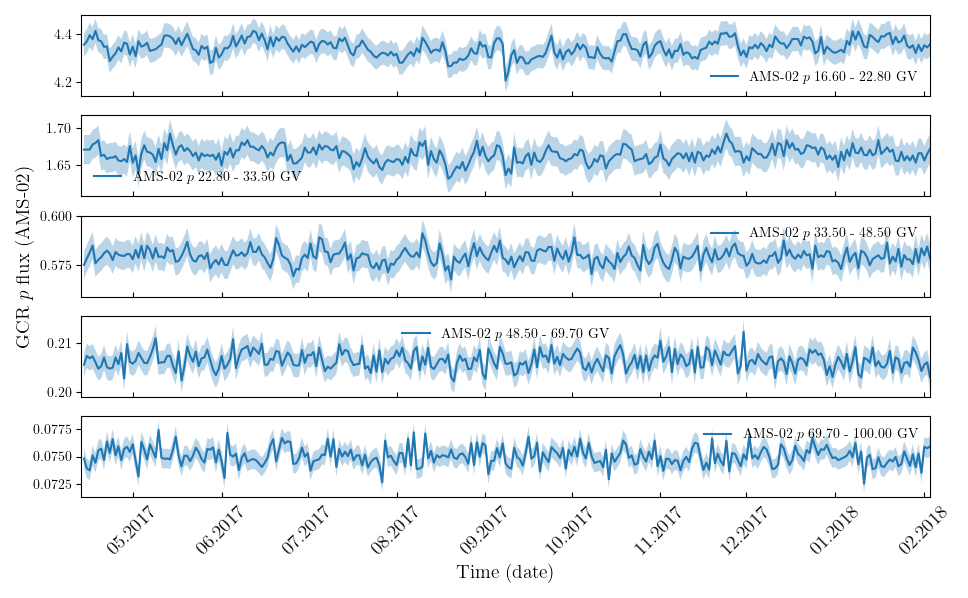}{0.80\textwidth}{(b)}}
\caption{Daily galactic proton observations measured by AMS-02 for the 13 April 2017 - 03 February 2018 (BR 2506 - 2516) as reported by \cite{2021PhRvL.127A1102A}, over (a) the first five rigidity bins, 1.0 - 1.16 GV, 1.16 - 1.33 GV, 1.33 - 1.51 GV, 1.51 - 1.71 GV \& 1.71 - 1.92 GV, and  (b) the last five rigidity bins 16.60 - 22.80 GV, 22.80 - 33.50 GV, 33.50 - 48.50 GV, 48.50 - 69.70 GV \& 69.70 -100.0 GV. The shading indicates the total error in GCR proton flux as reported.  
\label{fig2}}
\end{figure*}

The presence of a large depression in the proton flux lasting for almost half a year, while solar minimum is being approached, is an unusual signature that is too large to be explained by any individual solar-eruptive event. The purpose of this work is to determine the cause of this depression and we do this by investigating two approaches. First, we consider a possible large-scale change to the HMF due to a change in the behaviour of the Sun's global magnetic field. Secondly, we consider the effect of a combination of different solar activity events that potentially act together to produce a larger effect than any individual solar-eruptive event. The rest of the paper is outlined as follows: our two approaches of investigation are described in detail. We then perform a detailed investigation of observational data related to each approach. Based on the observational evidence, we evaluate both approaches, and the paper concludes with a discussion of the results. 

\begin{figure*}
\gridline{\fig{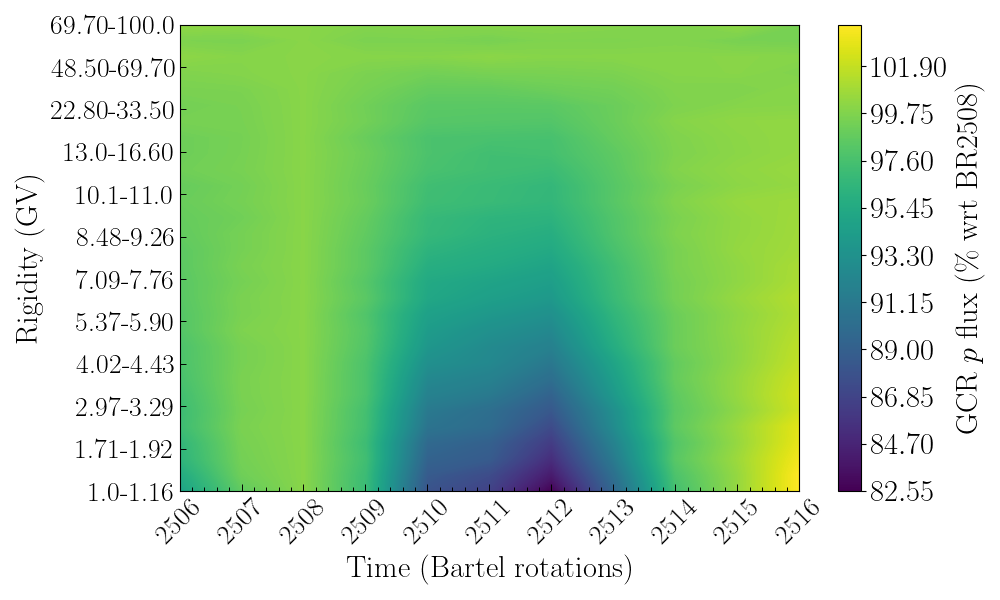}{0.53\textwidth}{(a)}
       \fig{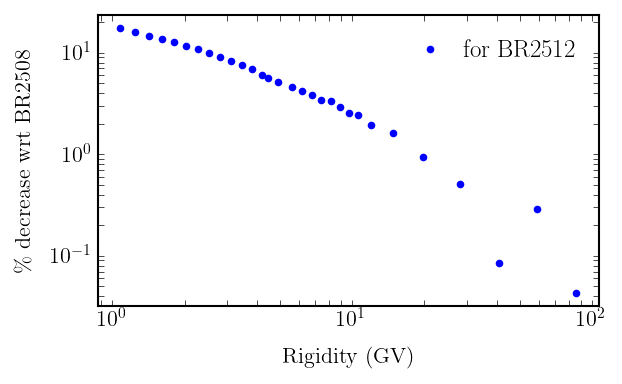}{0.45\textwidth}{(b)}}
\caption{(a) Normalized  BR-averaged GCR proton flux variation from BR2506 to BR2516 over the rigidity bins 1.0 -1.16 GV to 69.7-100.0 GV. Flux is normalized with respect to the BR2508. (b) The \% decrease in flux at BR2512 with respect to BR2508 over rigidity bins up to 100.0GV.       
\label{fig3}}
\end{figure*}
\section{Possibilities: Long-term or Transient structures} \label{sec:Posiblity}

As stated above, the observed depression in Figure \ref{fig1} is too large to be attributed to any individual solar-eruptive event. Such a statement presupposes that the Sun is the origin of the depression, but this needs to be confirmed. To do this, we consider the daily proton flux from AMS-02 \citep{2021PhRvL.127A1102A}, for the period in question, over a range of rigidity bins. If the main cause were due to some extra-heliospheric phenomenon, it would be expected that similar signatures would be observed in both high and low rigidity bins. If the origin of the depression were solar, however, then only the lower rigidities would be expected to carry the signature. Figure \ref{fig2} displays the daily proton fluxes from AMS-02 for (a) low and (b) high rigidity bins. By comparing the panels in Figure \ref{fig2} (a) and (b), it is clear that the consistent signature for low rigidities is lost at higher rigidities. This result is also conveyed in Figure \ref{fig3} (a), which displays a heat map for the normalized proton depression for many rigidity bins across the Bartel rotations covering the event. Figure \ref{fig3} (b) shows that the normalized decrease varies across rigidity for Bartel rotation 2512, when the minimum occurred.  These figures indicate that the signature of the depression begins to weaken for rigidities $\gtrsim 30$ GV, which is in line with values quoted in the literature \citep{2017AdSpR..60..848P}. Thus, we attribute a solar origin to the GCR depression.

With the data indicating a solar origin, two possible explanations for the origin of the depression need to be considered. The first is that, rather than solar activity being the primary cause, medium to long-term modulation effects, such as the observed depression, are functions of changes in the Sun's global magnetic field. Such an effect is referred to as a \emph{minicycle} \citep{1977ApJ...213..263G, 2000JGR...10518315W}, and is the first possibility for the observed depression that we will discuss. The second possibility that we consider is that the source of the observed depression lies primarily in solar activity. By \emph{activity}, we refer primarily to \emph{coronal mass ejections} (CMEs) and \emph{interaction regions} (IRs), that will be described in more detail later. We will investigate if a combination of CMEs and IRs can create the observed depression. We now describe each of the two possibilities in more detail.

\subsection{The Minicycle}

In 1974, a year-long modulation event occurred, near to solar minimum, that was qualitatively similar to the half-year-long observed depression of this study \citep{1977ApJ...213..263G, 1993JGR....9815231C,2000JGR...10518315W,2001SSRv...97..373W}. This event has been named the \emph{1974 minicycle} and it has been argued that its origin is due to a change in the Sun's global magnetic field and not transient solar-eruptive effects \citep{1995JGR...10021717T, 2000JGR...10518315W}. There are several key characteristics that define the 1974 minicycle. The observed depression for this event, detected in Mount Wellington NM data, mirrors very closely an increase in the magnitude $B$ of the HMF. This period is also matched closely by an increase in the tilt angle of the solar dipole, with a maximal change of almost 30$^\circ$ during this period. \cite{2000JGR...10518315W} argue that these changes in the solar magnetic field are the cause of the observed modulation event due to the absence of any enhanced CME activity. With its qualitative similarity to the observed depression of 2017, we test the minicycle hypothesis as a possible source of the observed depression. To do this, we will investigate the behaviour of $B$ and the tilt angle and evaluate their effects relative to any strong solar activity.

\subsection{CMEs and IRs}
Amongst the plethora of solar-eruptive phenomena, it is the large-scale structures of CMEs and IRs that are considered to play the most prominent roles in GCR modulation \citep{2000SSRv...93...55C, 2020ApJ...904....3M, 2020SoPh..295..104D, 2022A&A...658A.187D, 2024arXiv230811926G}. CMEs originate in the solar atmosphere and take the form of magnetic flux ropes that are ejected into space. The expansion of these ropes (also referred to as magnetic clouds) with increasing radial distance from the Sun provides a `magnetic obstacle' for low-rigidity GCRs. The classical signature of CME-induced modulation is the \emph{Forbush Decrease} (FD) \cite{1937PhRv...51.1108F}. Although the shape of a FD is qualitatively similar to the BR-averaged 2017 observed depression, the duration of a FD is of the order of a few days and not half a year. Thus, not one but many CMEs would be required to produce a combined effect long enough to span the observed dip. The merging of several CMEs beyond the Earth, forming Global Merged Interaction Regions (GMIRs), has long been studied \citep{1993JGR....98....1B}. At these large distances, the CMEs have expanded to become much larger than at 1 AU, thus having a form to create a longer-lasting modulation event. At 1 AU, it remains to be seen if a sequence of CMEs (near solar minimum) is enough to produce a GMIR-like effect that is sufficient enough to create the observed dip.  

In addition to CMEs, there is also the presence of IRs. IRs are formed when the fast solar wind, from coronal holes, catches up with the ambient slow solar wind. The result is a compression of plasma in this interacting region followed by a rarefaction in the fast solar wind \citep{1999SSRv...89...21G}. Such events are referred to as \emph{stream interaction regions} (SIRs). Due to the longevity of some coronal holes, SIRs may persist for several rotations. In this case, they are referred to as \emph{corotating interaction regions} (CIRs) in the literature. CIRs are know to produce FD-like depressions, which can be longer-lived than those due to CMEs but are typically weaker in magnitude \citep{2020ApJ...904....3M, 2022A&A...658A.187D, 2024A&A..00000R, 2024arXiv230811926G}. 
In the context of modelling the features of CIRs on GCR modulation, see e.g. \cite{2017ApJ...837...37K}; \cite{2020ApJ...899...90L,2024ApJ...961...21L}.

In order to determine whether or not solar activity is the primary cause of the observed depression, a record of CMEs and IRs during this period needs to be made. The contributions of these events then need to be identified in the GCR proton flux time series to confirm that the combined effects of individual events lead to the prolonged existence of the observed depression.

\subsection{Possibility of a minicycle}
\begin{figure}[ht!]
\centering
\begin{tikzpicture}
\node (t0 at (0,0){\includegraphics[width=0.9\textwidth]{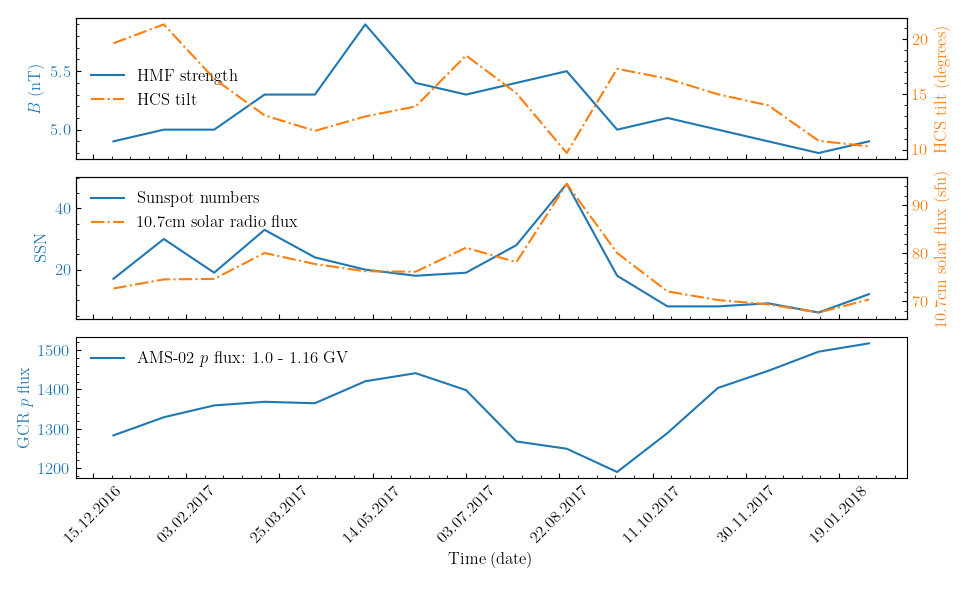}};
    \node at(3.65,4.25){(a)};
    \node at(3.65,1.7){(b)};
    \node at(3.65,-1.0){(c)};
\end{tikzpicture}
\caption{Time variation over a BR resolution of, top: IMF strength (blue) \&  HCS tilt angle (orange), middle: sunspot numbers (blue) \& 10.7cm solar radio flux (orange) and bottom: GCR proton flux in the rigidity bin 1.0-1.16 GV, for 26 December 2016 to 03 February 2018 period.} 
\label{fig4}
\end{figure}
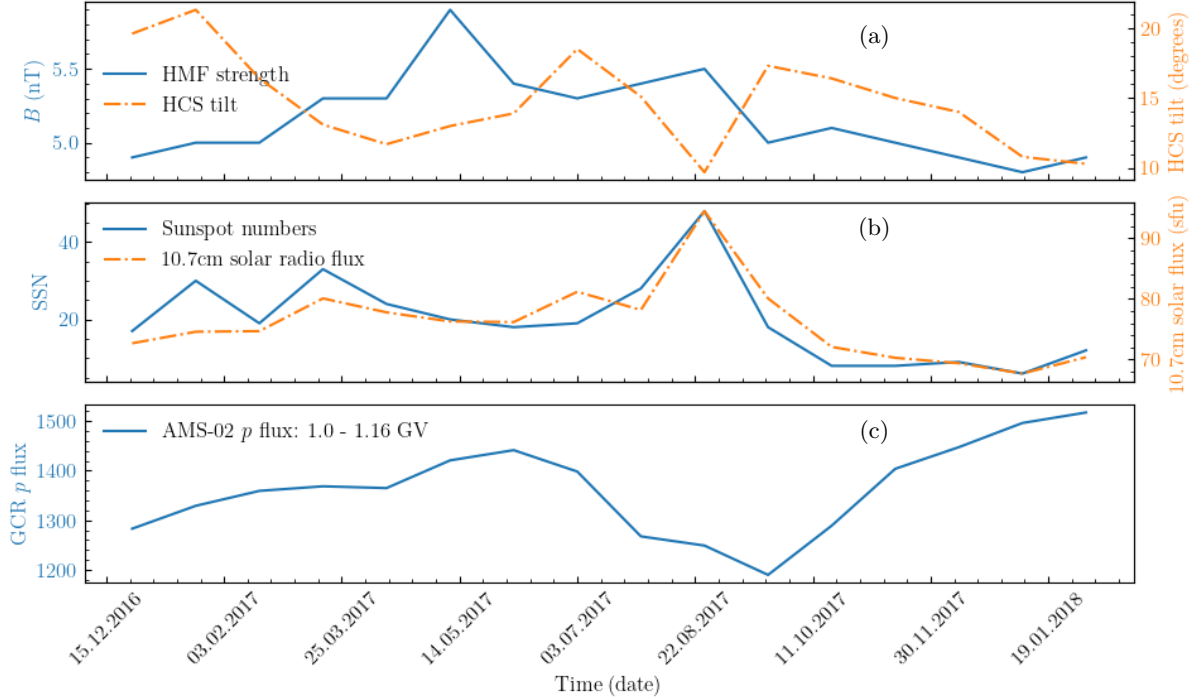

In order to test the possibility of a minicycle, we must seek the key characteristics reported by \cite{2000JGR...10518315W} for the 1974 minicycle. The main observations are an increase in the HMF field strength $B$ with a co-temporal change in the tilt angle of the solar dipole. Further, according to the description of the 1974 minicycle, both of these signatures must occur without an enhancement of CME activity. For the 1974 minicycle, $B$ increased by $\sim$1.5 nT and the tilt angle by $\sim30^\circ$. For the period of the observed depression under study, the key observations are displayed in Figure \ref{fig4}.

In Figure \ref{fig4} (a), the time variations, on a BR resolution, for the HMF magnitude $B$ (obtained from the Advanced Composition Explorer \citep{1998SSRv...86....1S} through the \href{https://omniweb.gsfc.nasa.gov/}{OMNIWeb Plus Service}) and the heliospheric current sheet (HCS) tilt angle (obtained from the \href{http://wso.stanford.edu/Tilts.html}{Wilcox Solar Observatory}) are displayed. To compare with these trends, the sunspot number (SSN), obtained through the \href{https://omniweb.gsfc.nasa.gov/}{OMNIWeb Plus Service} \citep{2000SSRv...93...11S},  and $F_{10.7}$ radio flux (obtained from the \href{https://www.spaceweather.gc.ca/forecast-prevision/solar-solaire/solarflux/sx-5-flux-en.php}{Dominion Radio Astrophysical Observatory}) are displayed in Figure \ref{fig4} (b), and the proton  flux from AMS-02, for the rigidity bin 1.0-1.16 GV, is shown in Figure \ref{fig4} (c). From the period starting at the end of 2016 and ending near the start of May 2017, there is an increase in $B$ of $\sim1$ nT and a decrease in the HCS tilt angle of $\sim10^\circ$. Although smaller than the changes for the 1974 minicycle, these values are indicative of the possibility of a minicycle. Further corroboration be found in Figure \ref{fig4} (b) by comparing the proxies of solar activity, SSN and F$_{10.7}$, for this period. There is little variation in both quantities, particularly during the main peak in $B$ in May 2017, and this is suggestive of no enhancement in solar eruptive activity. Further supportive evidence can be found by looking at the flares in this period. In Appendix \ref{appendix}, a list of all active regions from the end of 2016 until the start of 2018 (covering the period of the graphs in Figure \ref{fig4}) is presented. From the end of 2016 until the start of June 2017, there were only eight M-class flares (all pertaining to AR12644). All other flares during this period were C-class and below. Since the M-class flares occurred in the period from March 27th to April 5th 2017, it is unlikely that this activity was responsible for the peak in $B$, which occurred almost one month later. 

Until now, the evidence has been pointing in the direction of a possible minicycle. However, upon inspection of Figure \ref{fig4} (c),  the peak in $B$ and the local minimum in the HCS tilt angle are approximately co-temporal with a local maximum in the proton flux. This observation differs from that of the 1974 minicycle, in which there were already significant dips in proton and NM data when $B$ was at its maximum. It is only by the start of June 2017, after the peak activities of $B$ and the HCS tilt angle, that we find the observed depression in the GCR data. This behaviour corresponds to another increase in $B$ and decrease in the HCS tilt angle. The situation, however, is different from before. There are now large peaks in the SSN and F$_{10.7}$ data co-temporal with the peak in $B$ and the minimum in the HCS tilt. This information suggests that there is an increase in solar eruptive activity contributing to both the changes in $B$, the HCS tilt angle and the proton flux. From mid-August to mid-September 2017, there was a group of highly eruptive active regions (AR12671 - AR12674) that were responsible for the peaks in Figure \ref{fig4} (b). One of these regions, AR12673, also produced four X-class flares, including the strongest flare of the solar cycle \citep{2018A&A...619A.100H, 2020MNRAS.494..975L,2023ApJ...958..175R}.

To summarize, while indicative minicycle signatures appeared before the observed depression, it is clear that significant enhanced eruptive activity, co-temporal with the observed depression, cannot be excluded. For this reason, we cannot attribute the observed depression to a minicycle, at least in the sense of the 1974 minicycle described by \cite{2000JGR...10518315W}. Instead, to fully reveal the origin of the observed depression, we must go to our second possibility and examine the combined effects of solar activity.

\section{Solar activity} \label{sec:space weather}
\subsection{General picture}
\begin{figure}[h]
    \centering
    \begin{tikzpicture}
        \node at (-6.5,3) {(a)};
    
        \draw[thick,dashed] (-6,-3)--(-6,3);
        \node at (-6,-3.3) {$C(n)$};
        \draw[thick,dashed] (-4,-3)--(-4,3);
        \node at (-4,-3.3) {$C(n+1)$};
        \draw[thick,dashed] (-2,-3)--(-2,3);
        \node at (-2,-3.3) {$C(n+2)$};

         \draw[domain=-6:-4, smooth, variable=\x, blue,thick] plot ({\x}, {(\x+6)*(\x+4.5)});

         \draw[domain=-4:-2, smooth, variable=\x, blue,thick] plot ({\x}, {(\x+4)*(\x+2.5)+1});

        \node at (1.5,3) {(b)};
        \draw[thick,dashed] (6,-3)--(6,3);
        \draw[thick,dashed] (4,-3)--(4,3);
        \draw[thick,dashed] (2,-3)--(2,3);

        \draw[domain=2:2.5, smooth, variable=\x, blue,thick] plot ({\x}, {(\x-2)*(\x-3.5)});

        \draw[domain=2.5:4, smooth, variable=\x, red,thick] plot ({\x}, {(\x-2.5)*(\x-4.1)*(\x-0.5)-0.5});

        \draw[domain=4:6, smooth, variable=\x, blue,thick] plot ({\x}, {(\x-4)*(\x-5.5)-1.025});

        \node at (6,-3.3) {$C(n+2)$};
        \node at (4,-3.3) {$C(n+1)$};
        \node at (2,-3.3) {$C(n)$};
        
    \end{tikzpicture}
    \caption{Idealized representations of GCR proton flux, approaching solar minimum, over two Carrington rotations, with the $n$th dashed line marking the start of Carrington rotation $C(n)$. (a) shows the expected behaviour due only to CIRs. (b) shows the expected behaviour due to a disturbance, e.g. a CME, during the first Carrington rotation.}
    \label{fig:gen_pic}
\end{figure}
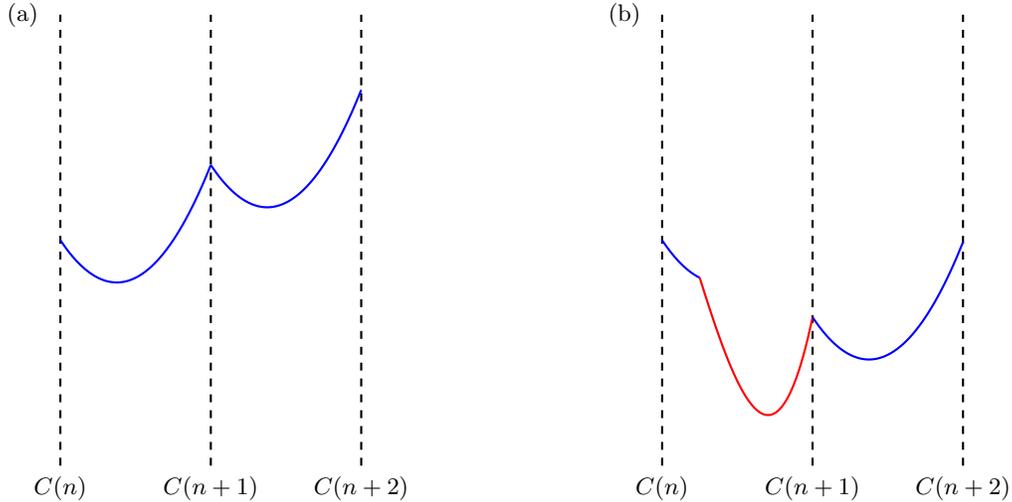

To investigate how solar activity may be the cause of the observed depression, it is important to identify all the key elements (CMEs, SIRs and CIRs) that contribute to it. Before performing this in detail, it is useful to present a simplified picture of what general behaviour is expected. Such a picture is presented in Figure \ref{fig:gen_pic}.

Figure \ref{fig:gen_pic} (a) displays the idealized behaviour of the GCR proton flux over two Carrington rotations as the solar cycle is approaching solar minimum. The general trend, averaged over Carrington rotations, is an increase in the proton flux. However, the ``full resolution'' rise is not monotonic due to solar rotation. $C(n)$ represents the $n$th Carrington rotation, with subsequent rotations being labelled $C(n+1)$, $C(n+2)$, etc. Within each Carrington rotation, the dip is caused by coronal holes producing CIRs. From the point of view of the impact on GCR flux, the impact of a CIR follows the rotation. Thus, instead of discussing CIRs per se, we will consider Carringtion rotations as the markers of CIR impact on GCR flux. Such a choice is clear from the picture in Figure \ref{fig:gen_pic} (a), but it will also become clear when examining the proton flux data in detail. 

In Figure \ref{fig:gen_pic} (b), there is a disturbance, shown in red, between $C(n)$ and $C(n+1)$. This disturbance represents the possible effect of a CME or a combination of CMEs and SIRs. In this picture, the disturbance ends at $C(n+1)$ and the proton flux in the next Carrington rotation proceeds as in Figure \ref{fig:gen_pic} (a). The effect of the disturbance has been to reduce the level of the proton flux at $C(n+2)$ compared to its ``undisturbed'' value in (a). If many such disturbances occur, they may delay the rise of the proton flux and create the signatures displayed in Figures \ref{fig1} and \ref{fig3}. We now investigate this possibility in the data.

\subsection{Data analysis}
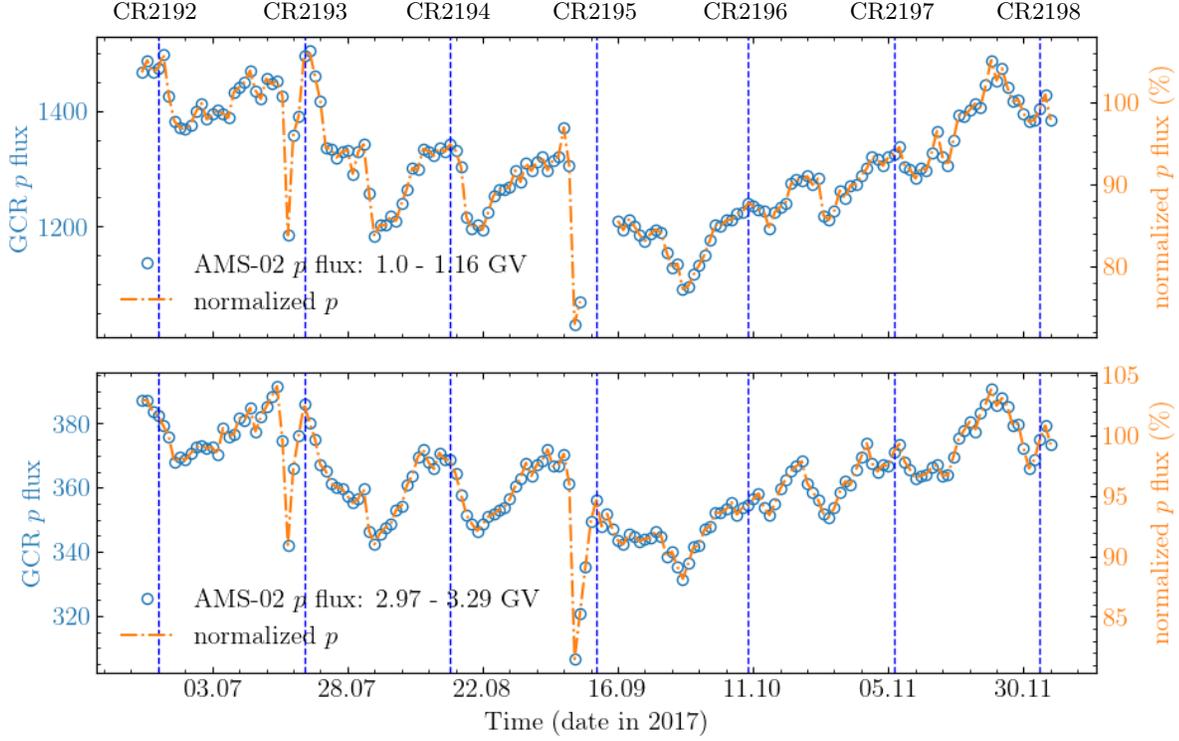
\begin{figure}[ht!]
\begin{tikzpicture}
\node (t0) at (0,0){\includegraphics[width=0.9\textwidth]{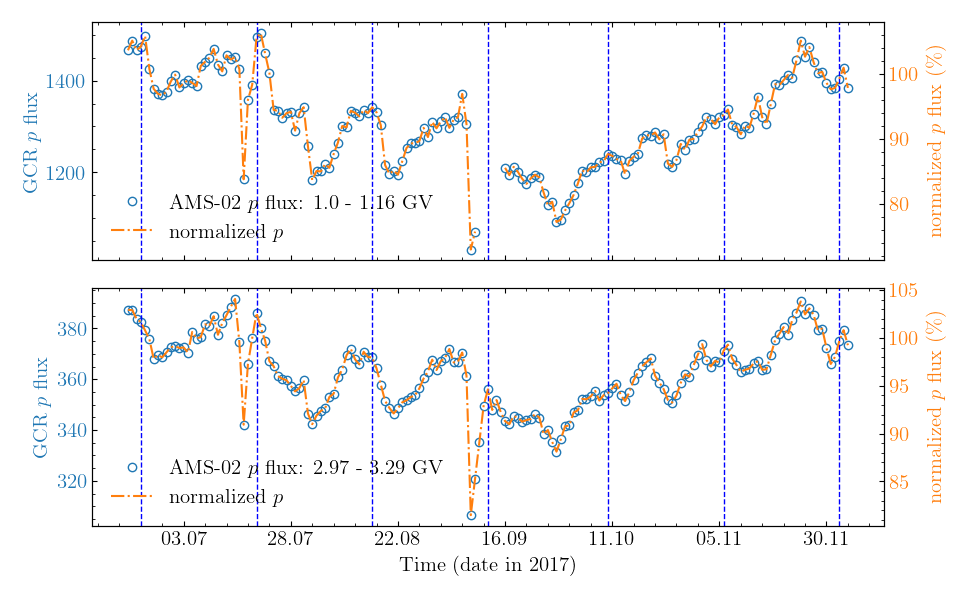}};
\node at (-5.8,5){CR2192};
\node at (-3.8,5){CR2193};
\node at (-1.9,5){CR2194};
\node at (0.05,5){CR2195};
\node at (2.05,5){CR2196};
\node at (4,5){CR2197};
\node at (5.95,5){CR2198};
\end{tikzpicture}
\caption{AMS-02 reported daily resolution GCR protons from 20 June to 05 December, 2017 at 1.0 - 1.16 GV (upper panel) and 2.97 - 3.29 GV (lower panel). The left scale is proton flux and right scale is normalized flux with respect to 06 June, 2017. The vertical blue lines indicate the beginnings of the Carrington rotations during this period. During the 09 - 15 September, 2017, daily proton AMS-02 measurements were not available below the 2.97- 3.29 GV rigidity bin.} 
\label{fig_ams_daily_dip}
\end{figure}

Figure \ref{fig_ams_daily_dip} displays the daily proton fluxes at low rigidities (bins 1 - 1.16 GV and 2.97 - 3.29 GV) over the period of the depression from 23 June to 03 December 2017. To aid the interpretation of the results, the beginnings of Carrington rotations are indicated by blue vertical lines and are labelled above.

This figure will provide a reference point for the description of solar activity. In addition to this figure, Table \ref{tab1} displays the start and end times of the Carrington rotations (CIRs) and major SIRs and CMEs relevant for the period. The strengths of the flares associated with each CME are also displayed, providing a relative indicator of the amount of energy released, together with their positions of origin on the solar surface and associated active region numbers.  Finally, the $\%$ decrease of proton flux associated with each event, as seen in Figure \ref{fig_ams_daily_dip}, is displayed. 

All of the data for Table \ref{tab1} have been collected from online catalogues for a range of instruments. For data of CME arrival times at 1 AU, there are the resources of \href{https://izw1.caltech.edu/ACE/ASC/DATA/level3/icmetable2.html}{Richardson \& Cane} \citep{2003JGRA..108.1156C,2010SoPh..264..189R}, \href{https://helioforecast.space/icmecat}{HELIO4CAST} \citep{2020ApJ...903...92M},\href{https://space.ustc.edu.cn/dreams/wind_icmes/}{WIND data from DREAMS} \citep{2016SoPh..291.2419C} and \href{https://stereo-ssc.nascom.nasa.gov/data/ins_data/impact/level3/LanJian_STEREO_CME_List.txt}{STEREO} \citep{2018ApJ...855..114J}. Carrington rotation (CIR) times and labels can be obtained from the  \href{http://wso.stanford.edu/Tilts.html}{Wilcox Solar Observatory} and SIR times are available from \href{https://stereo-ssc.nascom.nasa.gov/data/ins_data/impact/level3/LanJian_STEREO_SIR_List.txt}{STEREO} \citep{2019SoPh..294...31J}. 
Flare strengths and active region locations on the solar surface can be obtained from \href{https://www.solarmonitor.org/}{SolarMonitor} \citep{2002SoPh..209..171G}.

\newpage
\begin{longtable}{l|l|l|l|l|p{0.90cm}}
\caption{Identified solar activity during the June - December (CR2192 - CR2197) 2017 period (times correspond to activity at 1 AU). The GCR \% decrease is for the rigidity bin 1.0-1.16 GV. Generative active regions and locations on the solar surface are displayed for CMEs, together with associated flare strengths.\label{tab1}} \\
\hline
 Start  & End &  Structure & Region, location & Flare strength & \% dec. \\ 
\hline
\textbf{23-06T00:00} & \textbf{20-07T07:00} & \textbf{CR2192} & -- & -- & 9.25 \\
 23-06T03:00 & 25-06T22:00 & SIR & -- & -- & -- \\
 01-07T00:00 & 02-07T00:00 & SIR & -- & -- & 2.02\\
 04-07T02:53 & 06-07T09:14 & CME & AR12664, N18W85& B & 0.83 \\
 04-07T04:00 & 06-07T04:00 & SIR & -- & -- & -- \\
 09-07T01:00 & 10-07T01:00 & SIR & -- & -- & 3.39 \\
 11-07T04:00 & 14-07T04:00 & SIR & -- & -- & -- \\
 15-07T12:00 & 16-07T12:00 & CME & AR12667, N11W68 & C8.4  & --\\
 16-07T05:10 & 17-07T20:00 & CME & AR12665, S09W33& M2.4 & 18.99 \\
 16-07T14:00 & 18-07T12:00 & SIR & -- & -- & -- \\
 19-07T19:00 & 21-07T22:00 & SIR & -- & -- & -- \\
 \textbf{20-07T07:00} & \textbf{16-08T13:00} & \textbf{CR2193} & -- & -- & 11.97 \\
 24-07T14:35 & 27-07T11:00 & CME & AR12665, S06W89& C1.3  & 1.10 \\
 25-07T05:43 & 27-07T10:49 & CME & --& B& 3.0 \\
 29-07T22:50 & 31-07T22:53 & CME & -- & B & 11.31  \\
 08-08T12:00 & 09-08T21:00 & SIR & -- & -- & --  \\
 \textbf{16-08T13:00} & \textbf{12-09T20:00} & \textbf{CR2194} & -- & -- & 10.56 \\
 17-08T00:00 & 21-08T14:00 & SIR & -- & -- & -- \\
 21-08T22:00 & 23-08T18:00 & CME & AR12671, N11E08& C & 0.70 \\
 23-08T01:00 & 25-08T20:00 & SIR & -- & -- & \\
 28-08T20:00 & 30-08T06:00 & SIR & -- & -- & 1.24 \\
 31-08T21:00 & 02-09T00:00 & CME & -- & -- & 1.06\\
 03-09T21:00 & 06-09T00:00 & SIR & -- & -- & 1.47 \\
 06-09T23:02 & 08-09T04:00 & CME & AR12673, S11W16& M5.5& -- \\
 07-09T22:28 & 10-09T21:00 & CME & AR12673, S08W33& X9.3 & 24.4  \\
 \textbf{12-09T20:00} & \textbf{10-10T03:00} & \textbf{CR2195} & -- & -- & -- \\
 13-09T16:50 & 14-09T18:00 & SIR & -- & -- & -- \\
 18-09T13:17 & 19-09T00:00 & SIR & -- & -- & --  \\
 19-09T02:56 & 21-09T09:30 & CME & AR12680, N08W25& B& 2.57\\
 25-09T14:00 & 29-09T12:00 & SIR & -- & -- & 4.40\\
 30-09T12:00 & 04-10T18:00 & SIR & -- & -- & -- \\
 \textbf{10-10T03:00} & \textbf{06-11T09:00} & \textbf{CR2196} & -- & -- & 3.18\\
 10-10T19:30 & 12-10T06:35 & SIR & -- &-- & -- \\
 15-10T07:49 & 15-10T22:47 & SIR & -- & -- & -- \\
 21-10T03:24 & 23-10T06:58 & CME & AR12684, N11W74 & B & 1.06 \\
 24-10T08:36 & 25-10T06:58 & CME & AR12685, S12E88 & M1.1& 5.04 \\
 28-10T12:00 & 31-10T20:00 & SIR & -- & -- & --\\
 05-11T07:00 & 06-11T16:00 & SIR & -- & -- & --\\
 \textbf{06-11T09:00} & \textbf{03-12T16:00} & \textbf{CR2197} & -- & -- & 3.95 \\
 11-11T20:00 & 15-11T16:00 & SIR & -- & -- & --\\
 14-11T01:08 & 14-11T14:13 & CME & AR12687, S08E52& B &  0.42\\
 15-11T11:47 & 15-11T19:23 & CME & AR12687, S08E52 & B & 4.22\\
 17-11T00:00 & 17-11T21:53 & SIR & -- & -- & --\\
 19-11T18:00 & 22-11T12:00 & SIR & -- & -- & --\\
 23-11T20:00 & 25-11T07:00 & SIR & -- & -- & 2.57\\
 01-12T18:00 & 03-12T18:00 & SIR & -- & -- & -- \\
\hline 
\end{longtable}


After the onset of the first Carrington rotation, CR2192, on 23 June, the proton flux reduced by almost 10\% before beginning to rise. This behaviour is qualitatively similar to that described in Figure \ref{fig:gen_pic} (a).
The rise is impeded, however, by the impact of SIRs and a CME associated with a weak B-class flare. Each of these structures causes relatively small depressions. The rise of the proton flux is halted suddenly close to the start of the next Carrington rotation by the impact of two stronger CMEs (with related C-class and M-class flares). The first, from AR12667, arrived on 16 July  and the second, from AR12665, on 17 July. The combined effect of these two CMEs was to produce a depression of almost 19\%. 

\begin{figure}[h!]
    \centering
    \begin{tikzpicture}
    \node (t0 at (0,0){\includegraphics[width=0.9\linewidth]{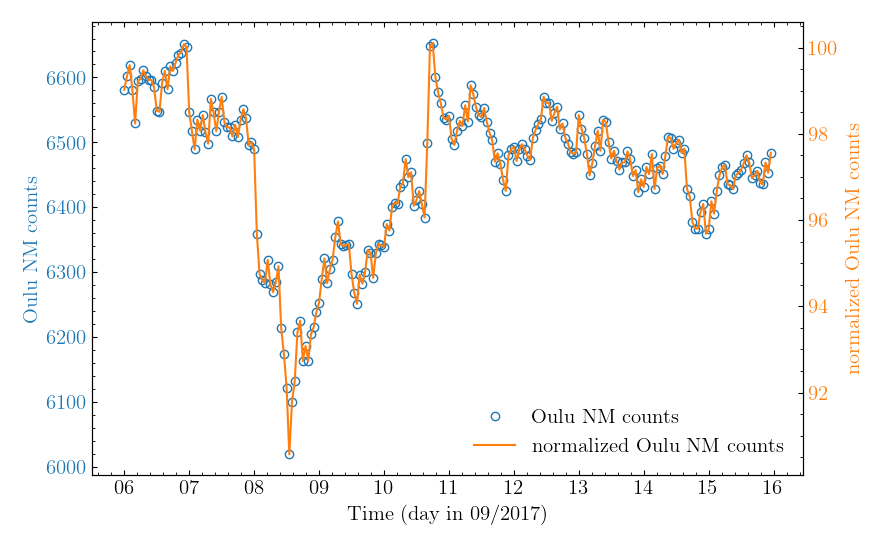}};
    \node at(-3.65,4.2){CME(M5.5)};
    \node at(-2.65,3){CME(X9.3)};
    \node at(2.4,3.5){CR2195};
    \node at(3.6,2.5){SIR};
    \end{tikzpicture}
    \caption{Oulu NM data covering the impact of the two strongest flares during the depression. Other features from Table \ref{tab1} are also indicated.}
    \label{fig:CME_double_impact}
\end{figure}
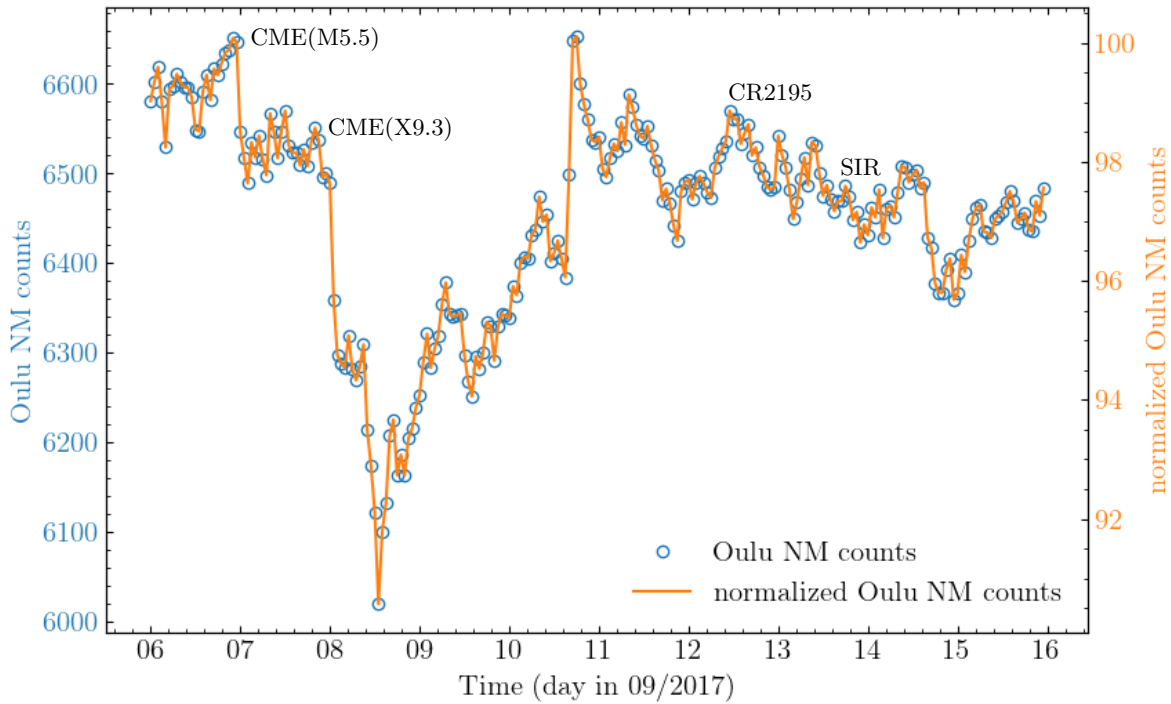

Although the proton flux at the start of CR2193 is at a similar level to that at the start of CR2192, its value is lower than it would have been had there not been the significant impact of the CMEs. The proton flux in the next two Carrington rotations behaves in a qualitatively similar manner compared to that of CR2193 - a CIR-related decrease at the start of the rotation followed by large impacts due to CMEs. The largest impact occurs near the end of CR2194 and is due to the combination of two CMEs associated with the strongest flares during this period. Indeed, the second of these CMEs is associated with the strongest flare of the entire solar cycle. Figure \ref{fig:CME_double_impact} displays the combined effect of these CMEs in higher-resolution NM data.

It is clear from Figure \ref{fig:CME_double_impact} that the recovery from the first CME impact is interrupted by the arrival of the second, which further prolongs the depression. Other features (CR2195 and an SIR) from available catalogues, that are listed in Table \ref{tab1}, are also labelled in this figure. Although there are CMEs in the remaining Carrington rotations of the depression, their relative impacts become smaller until, by CR2198, the proton flux begins to go ``back on track'' and follow the general rise expected as the solar cycle proceeds to its minimum. 

While the combination of various CMEs and SIRs in each Carrington rotation help to keep the proton flux lower than in ``quiet'' heliospheric conditions, the largest impacts come from CMEs associated with M-class and X-class flares. These CMEs are from AR12665, AR12673 and AR12685, and what is particular about them is that they all correspond to the same magnetic source, with different labels for different solar rotations. This particular magnetic source actually existed for five Carrington rotations and represents a localized region of the Sun in which complex magnetic field continued to emerge into the atmosphere and provide the energy required to produce the strong observed flares and CMEs. Thus, the main driver of this half-year depression in the proton flux has as its source, one highly-eruptive and evolving magnetic region (the most active of Solar Cycle 24). 

\subsection{Effect across rigidities}
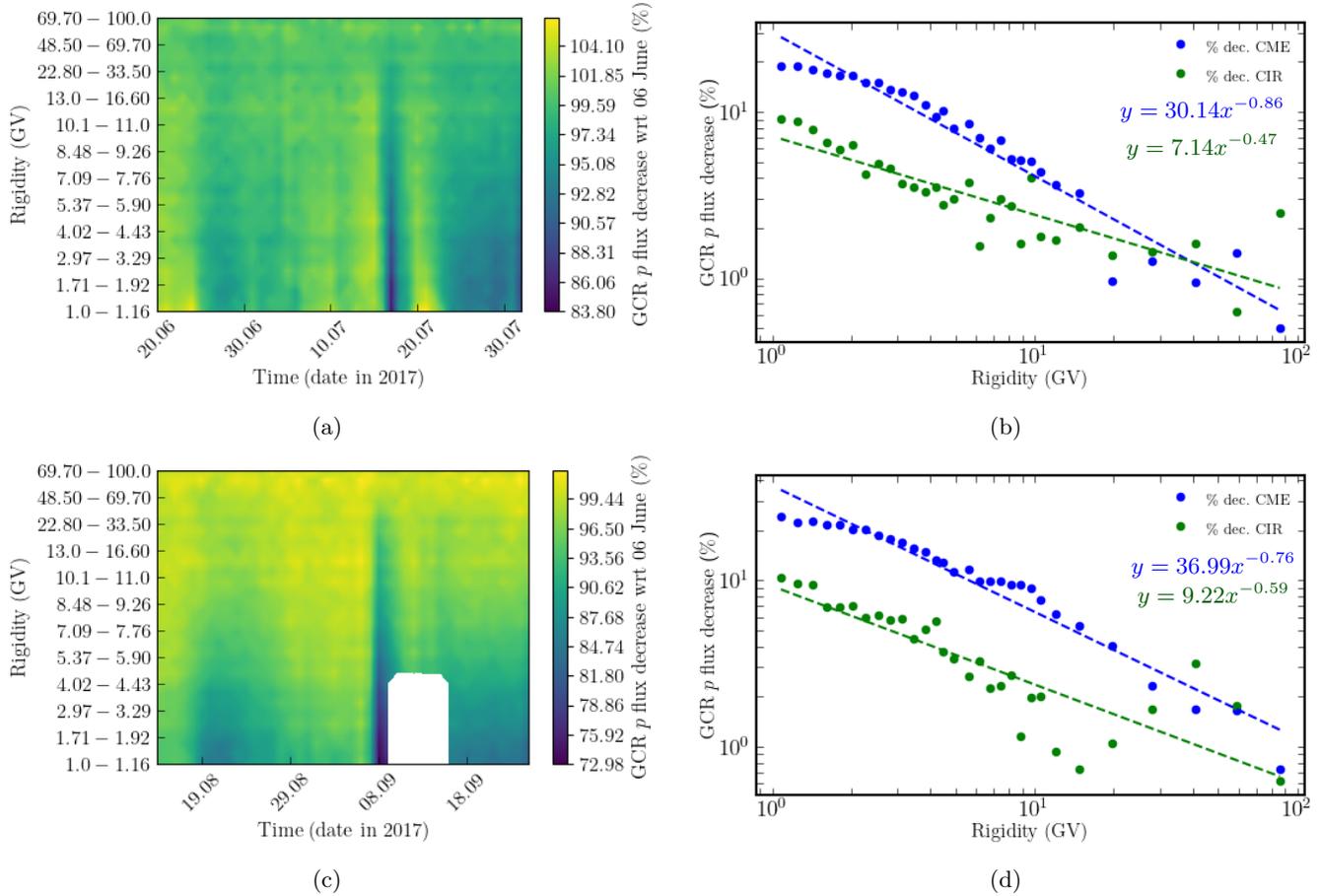
\begin{figure*}
    \centering
    \begin{minipage}[t]{0.50\textwidth}
        \includegraphics[width=\textwidth]{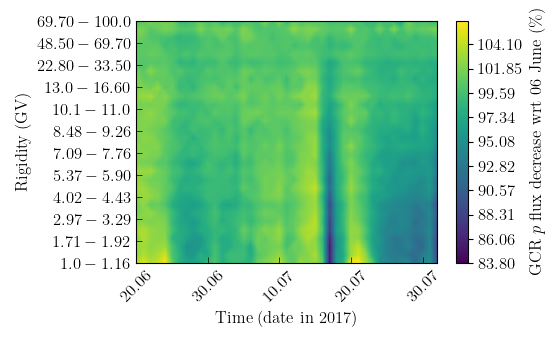}
        \centering (a)
    \end{minipage}
    \hfill
    \begin{minipage}[t]{0.485\textwidth}
        \includegraphics[width=\textwidth]{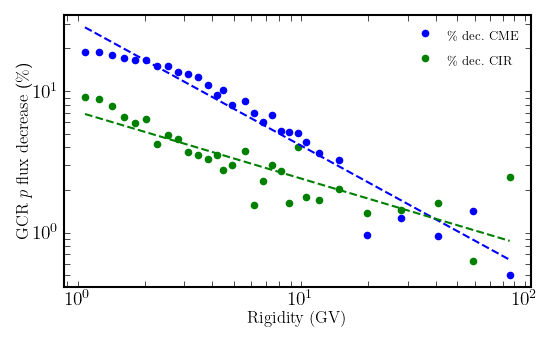}
        \centering (b)
    \end{minipage}
\definecolor{darkgreen}{rgb}{0.0, 0.35, 0.0}    
    \begin{tikzpicture}[overlay]
        \node at (7.35, 4.5) {\textcolor{blue}{$y=30.14x^{-0.86}$}};
        \node at (7.35, 4.0) {\textcolor{darkgreen}{$y=7.14x^{-0.47}$}};
    \end{tikzpicture}

    \begin{minipage}[t]{0.50\textwidth}
        \includegraphics[width=\textwidth]{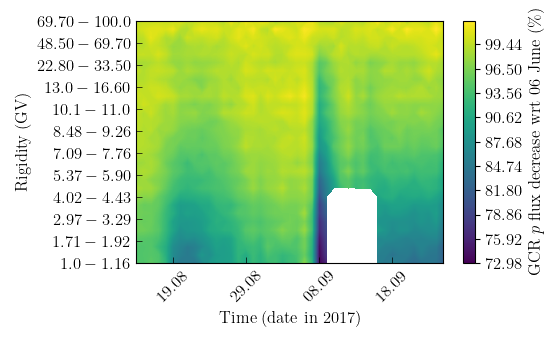}
        \centering (c)
    \end{minipage}
    \hfill
    \begin{minipage}[t]{0.485\textwidth}
        \includegraphics[width=\textwidth]{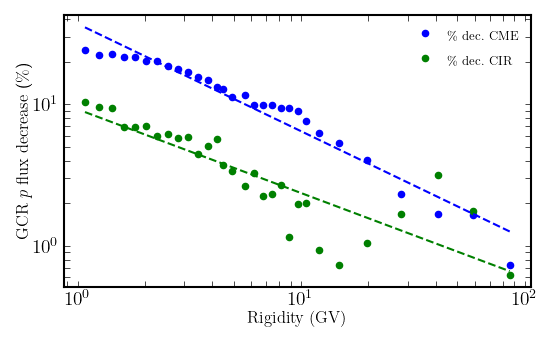}
        \centering (d)
    \end{minipage}

    \begin{tikzpicture}[overlay]
        \node at (7.5, 4.85) {\textcolor{blue}{$y=36.99x^{-0.76}$}};
        \node at (7.5, 4.45) {\textcolor{darkgreen}{$y=9.22x^{-0.59}$}};
    \end{tikzpicture}

\caption{Daily GCR proton flux reported by \cite{2021PhRvL.127A1102A}, normalized with respect to 06 June, 2017, over the whole reported rigidity range 1.0 - 100.0 GV, (a) for 20 June - 01 August, 2017 including CR2192 and (c) 14 August - 25 September, 2017 including CR2194. Note that for 09 - 15 September, below 2.97 - 3.29 GV, the proton data are missing.  Both heat maps reveal CIRs (thick and dark vertical bands), SIRs (weak, thin and dark vertical bands) and CMEs (strong, thin and dark vertical bands). (b) and (d) show the \% decrease as a function of rigidity for the first CIR and CME from (a) and (c) respectively. Power law fits are included to indicate the clear trends for lower rigidity particles. The formulae of the power laws are also indicated. 
\label{range_GV}}
\end{figure*}

So far, our focus has been on protons with low rigidities in the 1 - 1.16 GV and 2.97 - 3.29 GV bins. However, as indicated in Figure \ref{fig3}, the depression can still be detected clearly for rigidities up approximately 30 GV. We now show how individual impacts behave across different rigidities. Figure \ref{range_GV} displays information about how the \% decrease (normalized with respect to the value from 06 June) depends on rigidity for two periods covering 20 June to 01 August and 14 August to 25 September.

Figure \ref{range_GV} (a) shows a heat map of how the \% decrease behaves for rigidities in the range 1 - 100 GV over time. This period covers CR2192 and part of CR2193. The first strong CME on 17 July can be seen with a clear signature up to the 22.80 - 33.50 GV bin. The initial depressions at the start of Carrington rotations (CIRs) as well as SIRs are also clearly visible up to rigidities of around 30 GV. Figure \ref{range_GV} (b) displays the \% decrease during the depression of the first CIR shown in (a) and the main CME. This plot reveals that the \% decreases, for both the CME and the CIR, appear to follow power laws closely up to $\sim30$ GV. After this point, there is more scatter.

Figures \ref{range_GV} (c) and (d), covering CR2194 and part of CR2195, show similar properties to those in (a) and (b), despite the missing data from 09 to 15 September. Again, for this period, we see a clear power law-dependence of the \% decrease with rigidity up to $\sim30$ GV.

\section{Summary and discussion} \label{sec:summary}
In this work, we have investigated the source of an unexpectedly long depression in the GCR proton flux, reported by \cite{2021PhRvL.127A1102A}, lasting the second half of 2017. To achieve this task, we first identified the depression as being of solar origin. We then considered two possible sources for the depression - a minicycle due to a change in the global solar magnetic field, and a combined effect due to solar activity in the form of CIRs, SIRs and CMEs. After comparing the global properties of the HMF strength and the HCS tilt to the behaviour of the 1974 minicycle \citep{2000JGR...10518315W}, it was shown that the presence of some strong erupting active regions could not be ignored. Thus, in order to evaluate the how solar activity was connected to the depression, we identified all the catalogued CIRs and major SIRs and CMEs during this period. The proton flux depression lasted for five Carrington rotations, and we identified key features in each rotation. First, each Carrington rotation began, as expected, with an initial decrease in the proton flux. The subsequent rise was then impeded by a combination of CMEs and SIRs, and this general picture holds for each Carrington rotation although the relative decreases in the proton flux weakened by the final rotation. As part of this combination of different effects, four strong CMEs related to M-class or X-class flares, occurring in different Carrington rotations, caused the largest impacts. What is particular here is that all of these CMEs came from the same magnetic source, which perpetuated for five Carrington rotations. One phase of this magnetic source, when it was labelled AR12673, produced the strongest flare of the entire solar cycle \citep{2018A&A...619A.100H}. Therefore, from our findings, we conclude that the half-year depression observed in the GCR proton flux was caused by a combination of CIRs, SIRs and CMEs but was enhanced substantially due to strong CMEs emanating from an unusually active magnetic source (indeed, the most active of the cycle). We say ``unusually active'' because the phase of the solar cycle was approaching solar minimum. It is also for this reason, that the depression is more visible - it reverses the trend of the global proton flux rise (see Figure \ref{fig1}), whereas if the depression were in a phase approaching solar maximum, the depression would be in the same general direction as the global trend.

Although we excluded a minicycle as being the primary cause of the depression, this was based on the definition of the 1974 minicycle, for which there was no enhanced CME activity \citep{1995JGR...10021717T}. To our knowledge, within the field of solar dynamo theory, very little attention has been afforded to minicycles. Although in some works, minicycle-like features seem to appear \citep[e.g. Figure 4 of][]{2011GApFD.105..234B}, they have not been treated as an object of study in their own right. In our situation, we have some evidence of behaviour like the 1974 minicyle before the depression, but the difference is that there is a clear link between the depression and intense CME activity. However, we cannot exclude the possibility of a minicycle with a broader definition than the 1974 minicycle. Since all magnetic activity on the Sun ultimately derives from its dynamo, there must be some process that has led to the creation of the most eruptive active region as the solar cycle approaches its minimum. This may be another form of a minicycle which we suggest would be an interesting area of research in the future.

In this work, our main focus has been on the proton flux. In a follow-up to this work, we will model the modulation, \citep{2013SSRv..176..165P, 2014SoPh..289..391P,2021Physi...3.1190P} of protons and other GCRs, including helium \citep{2020Ap&SS.365..182N}, electrons and positrons \citep{2021ApJ...909..215A}, making use of AMS-02 data for these species \citep{2021PhRvL.127A1102A, 2022PhRvL.128w1102A, 2023PhRvL.130p1001A, 2023PhRvL.131o1002A}. This study will also help to explore the behaviour of particles with rigidities below 1 GV, for which observations are not currently available for the period under investigation. Further, we will also be able to simulate the modulation of anti-protons \citep[e.g.][]{2023ApJ...953..101A} for this period, and compare this to the modulation of protons in order to study charge-sign dependence for this particular type of event.

\begin{acknowledgments}
It is a pleasure to thank Paolo Zuccon for helpful discussions. O.P.M.A. and D.M. acknowledge support from a Leverhulme Trust grant (RPG-2023-182). D.M. also acknowledges support from a  Science and Technologies Facilities Council (STFC) grant (ST/Y001672/1) and a Personal Fellowship from the Royal Society of Edinburgh (ID: 4282). 

\end{acknowledgments}


\appendix

\section{List of solar active regions before and during the GCR depression}\label{appendix}

Here we present a list of active regions that appeared before during the depression, from 25 December 2016 to 03 February 2018. These details are needed in order to evaluate the minicylce and solar activity hypotheses discussed in this work. Beside the active region number, the corresponding Space-Weather Helioseismic and Magnetic Imager Active Region Patch \citep[SHARP;][]{hoeksema14} number is listed when available. The data of this table have been collated using the catalogues cited within the main text. In addition, the functionality of the ARTop code \citep{2023RASTI...2..398A} has been used for cross-checking details. The numbers of sunspots and sunspot area mentioned here are estimates based on the information from \href{http://helio.mssl.ucl.ac.uk/helio-vo/solar_activity/arstats/}{HELIO}.

\begin{longtable}{p{1.0cm}p{1.2cm}llll}
\caption{Active region and flare details covering the GCR depression. See the main text for details.} \\
\hline
 AR  & SHARP &  Initial time \& location & End time \& location & No. flares \& strengths & No. spots (area)\\
\hline
12619    & 6889 & 2016 Dec 25,  N03W79 & 2016 Dec 26, N03W91  & --           & -- (--) \\
12621    & 6895 & 2016 Dec 27, N11W57 & 2016 Dec 30, N10W91  & 8; B        & 04 (30)\\
12622    & -- & 2016 Dec 31, N12W71 & 2017 Jan 02, N11W91  & --            & 01 (10)\\
12623    & -- & 2017 Jan 02, S06W19 & 2017 Jan 07, S06W91  & --           & -- (--)\\
12624    & 6898 & 2017 Jan 04, S08W69 & 2017 Jan 06, S06W91  & 4; B        & 01 (--)\\
{12625}    & {6910} & 2017 Jan 13, N03E60 & 2017 Jan 24, N01W89  & 3; B        & 05 (80)\\
{12626}    & {6910} & 2017 Jan 14, N08E55 & 2017 Jan 26, N08W91  & 4; B        & 06 (160)\\
12627    & 6922 & 2017 Jan 21, N05E16 & 2017 Jan 29, N04W90  & 4, 1; B, C  & 14 (110)\\
{12628}    & {6920} & {2017 Jan 21, N12E37} & {2017 Jan 30, N11W91}  & {11, 6; B, C} & {9 (220)}\\
{12629}    & {6930} & {2017 Jan 25, N15E47} & {2017 Feb 05, N16W91 } & {22; B}       & {8 (190)}\\
{12630}    & {6937} & {2017 Jan 31, N16E42} & {2017 Feb 10, N16W91}  & --     & {1 (10)}\\
12631    & 6938 & 2017 Jan 31, S04W36 & 2017 Feb 10, N16W91  & 1; B        & 1 (10)\\
12632    & 6939 & 2017 Feb 01, N14W20 & 2017 Feb 07, N14W91  & 1; B        & 7 (60)\\
12633    & 6942 & 2017 Feb 03, N14W63 & 2017 Feb 05, N15W90  & --      & 3 (10)\\
{12634}    & {6946} & {2017 Feb 06, N02E58} & {2017 Feb 17, N03W91}  & {8; B}        & {5 (10)}\\
{12635}    & {6949} & {2017 Feb 10, N13E01} & {2017 Feb 17, N13W91}  & {15, 3; B, C} & {9 (110)}\\
12636    & 6950 & 2017 Feb 16, N11E41 & 2017 Feb 26, N11W91  & 3; B        & 4 (20)\\
12637    & 6951 & 2017 Feb 20, S03E49 & 2017 Mar 02, S04W91  & --       & 2 (20)\\
{12638}    & {6952} & {2017 Feb 21, N18E46} & {2017 Mar 04, N16W91}  & {20, 3; B, C} & {10 (150)}\\
12639    & 6958 & 2017 Feb 25, S09W46 & 2017 Mar 01, S08W91  & --       & 3 (20)\\
12640    & 6957 & 2017 Feb 27, N11E12 & 2017 Mar 07, N08W91  & 2; B        & 4 (30)\\
{{12641}}    & {{6961}} & {2017 Feb 28, N15E34} & {2017 Mar 09, N15W89}  & {19; B }      & {7 (100)}\\
{12642}    & {6961} & 2017 Mar 02, N14W01 & 2017 Mar 09, N15W91  & 4; B        & 5 (40)\\
12643    & 6967 & 2017 Mar 22, N09E60 & 2017 Apr 02, N08W91  & 2; B        & 2 (30)\\
{12644}    & {6972} & {2017 Mar 27, N12E18} & {2017 Apr 05, N13W91}  & {35,25,8; B, C, M} & {31 (570)}\\
{12645}    & {6975} & {2017 Mar 28, S09E47} & {2017 Apr 08, S09W91}  & {53,32; B, C} & {43 (700)}\\
12646    & 6974 & 2017 Mar 28, N06W72 & 2017 Mar 30, N06W91  & 6; B & 1 (20)\\
12647    & 6981 & 2017 Apr 01, N11W11 & 2017 Mar 30, N11W91  & --  & 4 (10)\\
12648    & 6982 & 2017 Apr 02, S03E60 & 2017 Apr 13, S03W91  & 2; B & 8 (50)\\
12649    & 6976 & 2017 Apr 04, N15W70 & 2017 Apr 06, N14W91  & --  & 4 (30)\\
12650    & 6983 & 2017 Apr 10, N08E59 & 2017 Apr 21, N08W91  & 16; B & 3 (40)\\
{12651}    & {6986} & 2017 Apr 19, N12E55 & 2017 May 01, N13W91  & 28, 3; B, C & 8 (150)\\
{12652}    & {6986} & 2017 Apr 21, N13E51 & 2017 May 01, N14W91  & 11, 1; B, C & 6 (30)\\
12653    & 6994 & 2017 Apr 22, S09E64 & 2017 May 04, S08W91  & 11; B & 2 (100)\\
12654    & 6999 & 2017 Apr 29, N10E40 & 2017 May 09, N11W91  & 7; B & 7 (90)\\
12655    & 7003 & 2017 May 05, N13E23 & 2017 May 14, N16W91  & 9; B & 6 (30)\\
12656    & 7013 & 2017 May 17, N12E48 & 2017 May 28, N11W91  & 1; B & 3 (30)\\
12657    & 7010 & 2017 May 17, N07E07 & 2017 May 24; N07W91  & 2; B & 2 (10)\\
{12658}    & {7015} & 2017 May 19, S08E24 & 2017 May 27, S07W91  & 1; B & 4 (20)\\
{12659}    & {7022} & {2017 May 22, N14E12} & {2017 May 30, N13W91}  & 31, 2; B, C & {13 (240)}\\
{12660}    & {7015} & 2017 May 23, S11W29 & 2017 May 29, S11W91  & 4; B & 10 (12)\\
{12661}    & {7034} & {2017 Jun 01, N06E65} & {2017 Jun 13, N07W91}  & {35, 16; B, C} & {13 (240)}\\
{12662}    & {7045} & {2017 Jun 14, N12E56} & {2017 Jun 25, N13W91}  & {--} & {3 (170)}\\
12663    & 7050 & 2017 Jun 16, N14W14 & 2017 Jun 22, N14W91  & 12; B & 7 (100)\\
12664    & 7058 & 2017 Jun 21, N18E58 & 2017 Jul 03, N18W91  & 12; B & 10 (130)\\
{12665}    & {7075} & {2017 Jul 06, S05E64} & {2017 Jul 18, S06W91}  & {79, 29, 2; B, C, M} & {32 (730)}\\
12666    & 7081 & 2017 Jul 13, N13W20 & 2017 Jul 19, N15W91  & 1; B & 9 (30)\\
12667    & -- & 2017 Jul 13, N12W91 & 2017 Jul 16, N12W91  & 7, 4; B, C & 1 (10)\\
12668    & 7090 & 2017 Jul 26, N03W42 & 2017 Jul 30, N03W91  &  -- & 2 (10)\\
12669    & 7097 & 2017 Jul 30, N18W39 & 2017 Aug 03, N18W91  & -- & 2 (10)\\
12670    & 7100 & 2017 Aug 03, S05E44 & 2017 Aug 14, S07W91  & 27; B & 3 (170)\\
{12671}    & {7107} & {2017 Aug 15, N11E59} & {2017 Aug 28, N13W91}  & {71, 22; B, C }& {31 (580)}\\
{12672}    & {7110} & {2017 Aug 21, N05E61} & {2017 Sep 02, N08W91}  & {29, 9, 1; B, C, M} & {19 (300)}\\
{12673}    & {7115} & {2017 Aug 30, S08E50} & {2017 Sep 10, S09W91}  & {2, 55, 27, 4; B, C, M, X }& {33 (1080)}\\
{{12674}}    & {{7117}} & {2017 Aug 30, N11E58} & {2017 Sep 12, N16W91 } & {36, 15; B, C }& {34 (980)}\\
12675    & 7120 & 2017 Sep 01, S07W36 & 2017 Sep 06, S07W91  & -- & 5 (50)\\
12676    & 7123 & 2017 Sep 03, S10W48 & 2017 Sep 07, S09W91  & -- & 6 (30)\\
12677    & 7122 & 2017 Sep 04, N19E53 & 2017 Sep 15, N17W91  & 1; B & 4 (20)\\
{12678}    & {7127} & {2017 Sep 06, N11E33} & {2017 Sep 15, N11W89 } & {2; C }& {6 (40)}\\
{12679}    & {7117} & 2017 Sep 08, N14W39 & 2017 Sep 12, N16W91  & 1; B & 4 (10)\\
{12680}    & {7131} & {2017 Sep 11, N09E54} & {2017 Sep 23, N07W91 } & { 8, 2; B, C }& {3 (190)}\\
12681    & 7144 & 2017 Sep 21, S12E61 & 2017 Oct 03, S13W91  & 11; B & 8 (120)\\
12682    & 7147 & 2017 Sep 25, S09E57 & 2017 Oct 07, S11W91  & 11; B & 7 (230)\\
{12683}    & {7148} & {2017 Sep 26, N11E62} & {2017 Oct 08, N13W91} & {39, 3; B, C }& {6 (440)}\\
12684    & 7164 & 2017 Oct 16, N11W46 & 2017 Oct 20, N11W91  & -- & 2 (10)\\
12685   & 7169 & 2017 Oct 22, S11E60 & 2017 Nov 02, S09W89  & 2, 1; B, M & 4 (70)\\
12686    & 7171 & 2017 Oct 24, N12E53 & 2017 Nov 04, N13W91  & 2; B & 3 (40)\\
12687    & 7189 & 2017 Nov 15, S08E52 & 2017 Nov 26, S09W91  & 10; B & 5 (90)\\
12688    & 7190 & 2017 Nov 18, N11W44 & 2017 Nov 22, N11W91  & 1; B & 2 (10)\\
12689    & 7192 & 2017 Nov 26, N13W20 & 2017 Dec 02, N13W91  & 6; B & 5 (60)\\
12690    & 7202 & 2017 Dec 07, N07W29 & 2017 Dec 12, N06W91  & 1; B & 3 (10)\\
12691    & 7203 & 2017 Dec 11, S03E30 & 2017 Dec 20, S03W91  & -- & 3 (10)\\
12692    & 7211 & 2017 Dec 21, N16E33 & 2017 Dec 31, N17W91  & 18; B & 12 (160)\\
12693    & 7222 & 2018 Jan 05, N18W52 & 2018 Jan 09, N20W91  & 1; B & 3 (20)\\
12694    & 7227 & 2018 Jan 09, S32W15 & 2018 Jan 15, S32W91  & -- & 3 (10)\\
12695    & 7228 & 2018 Jan 12, S08W71 & 2018 Jan 14, S09W91  & 1; B & 2 (10)\\
12696    & 7229 & 2018 Jan 16,  S11E13 & 2018 Jan 25, S12W91  & 7; B & 3 (30)\\
12697    & 7235 & 2018 Jan 31,  S10E36 & 2018 Feb 09, S09W91  & 3; B & 5 (10)\\
12698    & 7236 & 2018 Feb 03,  S09W07 & 2018 Feb 14, S03W91  & 1; B & 1 (10)\\
\hline
\label{TabA}
\end{longtable}

\end{CJK*}
\end{document}